# Time-domain THz spectroscopy reveals coupled protein-hydration dielectric response in solutions of native and fibrils of human lysozyme

Fabio Novelli[1], Saeideh Ostovar Pour[2], Jonathan Tollerud[1], Ashkan Roozbeh[1], Dominique R. T. Appadoo[3], Ewan W. Blanch[2], Jeffrey A. Davis[1,*]

[1] Centre for Quantum and Optical Science, Swinburne University of Technology, Victoria 3122, Australia

[2] School of Science, RMIT University, GPO Box 2476, Melbourne, Victoria 3001, Australia

[3] Australian Synchrotron, 800 Blackburn Road, Clayton, 3168 VIC, Australia

**ABSTRACT:** Here we reveal details of the interaction between human lysozyme proteins, both native and fibrils, and their water environment by intense terahertz time domain spectroscopy. With the aid of a rigorous dielectric model, we determine the amplitude and phase of the oscillating dipole induced by the THz field in the volume containing the protein and its hydration water. At low concentrations, the amplitude of this induced dipolar response decreases with increasing concentration. Beyond a certain threshold, marking the onset of the interactions between the extended hydration shells, the amplitude remains fixed but the phase of the induced dipolar response, which is initially in phase with the applied THz field, begins to change. The changes observed in the THz response reveal protein-protein interactions mediated by extended hydration layers, which may control fibril formation and may have an important role in chemical recognition phenomena.

Proteins are the most versatile macromolecules of the cell and the ability to maintain their native, functional state is fundamental to biological activity. A great number of mammalian disorders have some linkage with protein misfolding and aggregation. Amyloid fibril related disorders[1-4], for example, are associated with the unwanted filamentous aggregation of particular proteins or peptides. Human lysozyme is a good model system that has been used for decades to study amyloid fibrillogenisis. The common features of amyloidogenic proteins are β-sheet formation and fibrillar morphology[5], however, the detailed mechanism of β-sheet formation and the molecular level properties of fibrils are not yet understood despite being subjects of great interest[6-8].

The importance of the dielectric solvated environment, particularly of water molecules in mediating the secondary and tertiary structure of proteins, interactions between proteins[9,10], and conformational changes that lead to aggregation is widely appreciated[11-21], but poorly characterized.

Light at terahertz (THz) frequencies (0.1-10 THz or 3,000-30 μm) is strongly absorbed by water and provides a very sensitive probe of any perturbations to the hydrogen bonding network[22,23]. At the interface of a macromolecule in solution several things typically happen: the water molecules close to the surface will be strongly influenced and their dynamics retarded by charges on the surface of the macromolecule, leading to changes in the THz absorption spectrum[24,25]. The influence of protein-water interactions also extends beyond these tightly bound water molecules[26-29], causing changes to the hydrogen bonding network and thus to the contribution of both single water molecules and the larger domains to the THz absorption spectrum. Details of these changes, and how far these effects propagate from the surface of the macromolecule, remain open questions. Techniques based on NMR[30] and neutron or X-ray scattering[31] probe only the tightly bound water molecules, indicating that a sub-monolayer of water molecules is formed around each protein at near physiological conditions. Similar conclusions are drawn from terahertz spectroscopy when films[32-34] or relatively large protein concentrations are studied (≥ 60 mg/ml for hen egg white lysozyme[35]). However, when lower concentrations are used, terahertz techniques suggest that the population of hydration water is much larger and extends over multiple layers. Extended hydration shells, where the effect of the protein on the water molecules is evident, have been reported to vary between 15 Å and 25 Å from the solute surface for different proteins[25-29]. While these works are based on the simplistic assumption that the absorption coefficient of a complex system can always be written as the weighted sum of the absorption of its components[35], here we show how determining the complex dielectric response provides a much more informative and reliable assessment of the impact of protein-water interactions on the water environment. Another recent work investigated the full dielectric response of lysozyme films[34]. However, results on films[33,34] cannot be easily compared with results on solutions because additional collective modes appear upon film formation[36], and because films typically have much lower water content (compare e.g. the dielectric functions shown in Ref.[34] with the one of bare water in Ref.[23]).

Here we report high-resolution terahertz time-domain measurements of solutions containing native human lysozyme or fibrils down to concentrations as low as 5 mg/ml (corresponding to ~0.36% volume fraction) at room temperature. By determination of the complex dielectric function, we are then able to determine the amplitude and phase of the induced dipole in the volume containing the protein and the extended hydration water. Further details of the hydration layer and interactions between proteins are then revealed by tracking the concentration dependence of these values.

We measured the amplitude and the phase of the THz field transmitted through 0.5 mm thick lysozyme-water solutions at room temperature. We used large amplitude THz fields generated by tilted-front optical rectification[37-41] and detected by electro-optical sampling[42-44], as described in the Supplementary Information. The associated pulse fluence is lower than 2 $\mu J/cm^2$ and any thermal effects can be disregarded[45]. Human lysozyme was obtained from Sigma Aldrich (Australia). The samples were prepared in milli-Q water with pH value of 4.80 for solutions containing native proteins, and with pH 2.0 for solutions containing fibrils. In order to form fibrils of human lysozyme the samples were incubated at 96 °C for 16 hours in a solution with pH 2.0 [46]. The formation of fibrils following this protocol has previously been verified by AFM, TEM and Thiflavin T assay[46]. Additional details of sample preparation are discussed in the Supplementary Information. The distilled water and the protein-free solutions at pH 2.0 and 4.8 all have, within our resolution, the same optical absorption and refraction properties at the terahertz frequencies investigated in this work. In the following we simply take distilled water as the background for zero protein concentrations. The samples are well shaken before the measurements, and the fields transmitted by a reference and by the filled quartz cell are acquired alternately 10 times for a total acquisition window shorter than 15 minutes per sample.

In Figure 1 we report typical time-domain traces of the THz pulses transmitted by a reference (Figure 1a) and by 100 mg/ml solutions of native human lysozyme (blue curve in Figure 1b) and fibrils (red curve in Figure 1b). By comparing the blue and red curves in Figure 1b it can be seen that the terahertz response of fibrils and native solutions is distinctly different: the solution containing the fibrils displays lower transmission and is delayed in time. The inset of Figure 1b shows the spectral amplitude of the transmission, compared to the incident spectrum (the dips in the incident THz spectrum at and above ~ 1.1 THz are due to absorption by water molecules in the air[47]). The large drop (>20x) in the THz spectral amplitude transmitted by the protein solutions is primarily due to absorption by bulk water. However, by normalizing the transmission of the protein solutions to the transmission through pure water (see Supplementary Figure 1(a)), it can be seen that the spectral response of both native lysozyme and fibrils is flat and that the fibrils transmit less than the native lysozyme. These results are confirmed by additional measurements performed at the FAR-IR beamline of the Australian Synchrotron (see Supplementary Figure 1(b,c)).

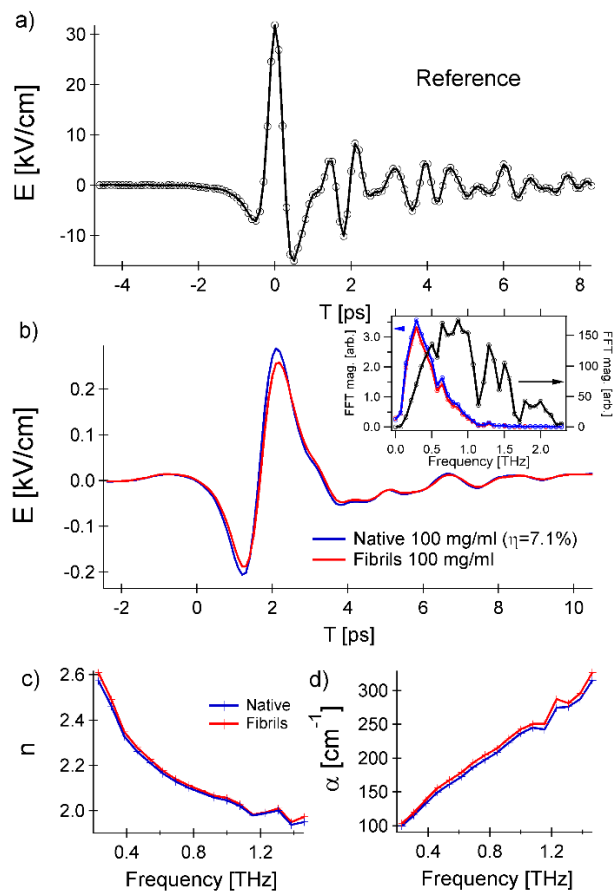

Figure 1. a) Amplitude of the pulsed THz field transmitted through the empty sample holder as a function of the sampling pulse delay. b) Typical THz pulses transmitted by water-based solutions containing 100 mg/ml native (blue) and fibrils (red). In the inset the magnitude of the Fourier transformations are shown as a function of frequency. c) The index of refraction for the lysozyme solutions and d) The corresponding absorption coefficient.

From the time-domain THz experiments we calculate the full dielectric properties of the solutions, both the index of refraction (Figure 1c) and the absorption coefficient (Figure 1d), assuming only the first Fresnel transmission coefficient is relevant (see Supplementary Information for details). It is important to note that scattering cannot explain the differences observed between fibril and native protein solutions (see Supplementary Information).

For the following analysis we focus on the average values for the optical properties over the 0.4-0.9 THz range. The average absorption coefficient is shown in Figure 2a for native (blue solid squares) and fibril (red solid squares) lysozyme solutions, together with the expected effect of removing a volume of bulk water equal to the protein volume (black line in Figure 2a). The optical properties depend on the volume fraction, $\eta$, of the proteins in solution. For native lysozyme we measured a variety of solutions with $\eta$ ranging from about 0.36% to 42.7% (or, equivalently, from 5 to 600 mg/ml, taking the volume of each protein as a simple sphere with radius ~ 15.9 A[48] and assuming ideal behavior [49]). Three distinct regions can be identified in Fig. 2(a) for the native human lysozyme (blue squares): above the solution saturation threshold of $\eta \sim 17.8\%$ (250 mg/ml) a strong deviation from water-removal effects is seen, as expected[35]; between $\eta = 7.1\%$ and 17.8% (100 and 250 mg/ml) a linear decrease is seen, showing reasonable agreement with bulk water-removal effects[35]; and at volume fractions in the range 0.36% – 3.6% (5-50 mg/ml) a flatter region is evident (zoomed in Figure 2b), which cannot be explained as removal of bulk water[29], but can be described in part by contributions from the water molecules that are affected by the presence of the protein[24,29]. These results are consistent with previous measurement of THz absorption of lysozyme over different parts of this concentration range[29,35].

The solutions containing fibrils of lysozyme have been studied below the saturation point, between $\eta \sim 0.36\%$ (5 mg/ml) and 14.2% (200 mg/ml). It is evident that the solutions with fibrils display a larger absorption and index of refraction compared to the native lysozyme (see Figure 1 and Figure 2).

In the following we focus on the results obtained below $\eta \sim 3.6\%$ (50 mg/ml) where the protein volume is known not to change for different concentrations[49].

To understand the volume-fraction dependent response of the fibril and native human lysozyme solutions shown in Figure 2 we begin by following the approach of Heyden et al.[24]. The starting point is to consider the protein volume as a void. The first order effect is simply the removal of bulk water, which leads to a reduction in the absorption linearly proportional to the volume fraction of the protein in solution (this is the solid black line in Figure 2a and Figure 2b). However, in a dielectric medium the applied field will also induce a polarization at the interface between the void and the dielectric. If we consider the protein to be a spherical void and water to be a static dielectric medium, Maxwell's equations tell us that the induced field will be in the opposite direction to the applied field, with magnitude given by $\boldsymbol{M}^M = -V\boldsymbol{P}\frac{3\epsilon_S}{2\epsilon_S+1}$ where $V$ is the void volume, $\boldsymbol{P}$ is the transverse bulk polarization, and $\epsilon_S$ is the dielectric constant of the solvent[24]. Under these assumptions the resultant average dipole moment induced over the solution, and hence the THz absorption, is further reduced.

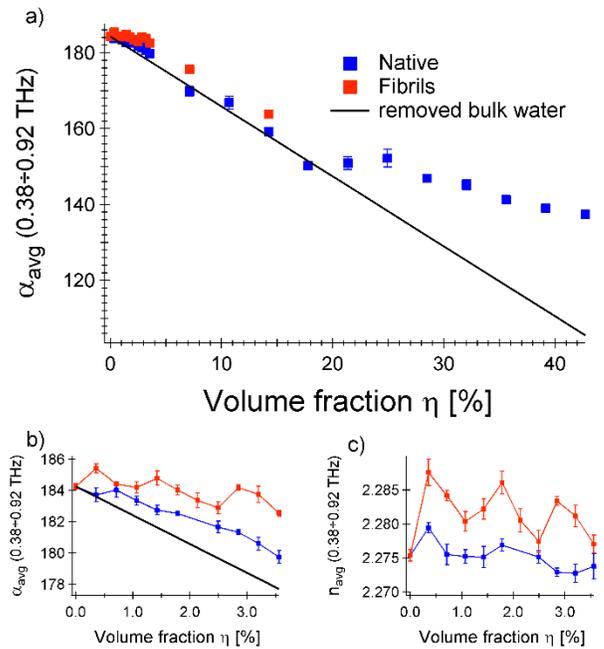

Figure 2. a) The average absorption coefficient in the 0.38-0.92 THz range of the lysozyme solutions is shown versus the volume fraction of lysozyme. b) Same as a) but on an expanded scale for small concentrations or volume fractions. Error bars are $\pm$ 1 standard deviations calculated from 10 measurements (see Supplementary Information). c) Average index of refraction at low concentrations.

Previous measurements of proteins (including lysozyme) in water solutions have, however, shown that rather than the THz absorption being lower than expected when considering only water removal, the THz absorption is actually higher[24-29], consistent with the results reported here (Fig. 2).

The absorption is enhanced because the protein is not a void and will typically have some absorption and surface charges, and because the water environment of the protein is not a static dielectric but is able to move and rotate to compensate for an electromagnetic perturbation. It is clear then that the pure Maxwell picture described above does not provide a complete description. Heyden et al.[24] thus defined the scalar quantity "$a$" relating the actual dipole moment induced in the volume of solution containing one protein and its hydration layers, $\boldsymbol{M}^{int}$, to the dipole moment predicted in the case of one spherical void in a static dielectric $\boldsymbol{M}^M$:

$$a = \hat{\boldsymbol{x}} \cdot \frac{\boldsymbol{M}^{int}}{\boldsymbol{M}^M} \qquad \text{Eq. 1}$$

where $\hat{\boldsymbol{x}}$ is an unitary vector along the direction of the applied electric field. This can be combined with the contribution from bulk water and its removal to give the complex, frequency-dependent dielectric response of the solution, $\epsilon(\omega)$:

$$\epsilon(\omega) = 1 + [\epsilon_S(\omega) - 1][1 - \eta] - \eta \left[ a(\omega) \frac{[\epsilon_S(\omega)-1]^2}{2\epsilon_S(\omega)+1} \right] \quad \text{Eq.2}$$

where $\eta$ is the volume fraction of the protein in solution and $\epsilon_S(\omega)$ is the dielectric function of the solvent alone[24]. In the case that $a = 1$ the induced dipole will follow the behaviour predicted from Maxwell's electrostatics $\boldsymbol{M}^{int} = \boldsymbol{M}^M$; for $a = 0$ the solvent rearrangement will perfectly compensate for the Maxwell dielectric field and the behaviour of the solution can be described simply as removal of bulk water.

In several previous works[28,50] the THz absorption of protein solutions decreases slower than expected for solvent removal, or in some cases even increases as a function of protein concentration[24-27,29]. This corresponds to a negative value of $a$ and an induced dipole, in the volume containing protein and hydration water molecules (i.e. $\boldsymbol{M}^{int}$), that is parallel with the applied field direction. This is consistent with the absorption behavior observed here; however, we also have the refractive index, which allows a reassessment of Eq.1 and the determination of both amplitude and phase of the induced dipole.

The analysis proposed by Heyden et al.[24], based on Eq. 1, indicates that only the component of the induced field in phase (negative $a$) or $\pi$ out of phase (positive $a$) with the applied field is relevant. In general, however, the induced dipole can have any phase relative to the applied field. Determination of the full complex dielectric function, as we obtain here, provides additional constraints that allow us to determine a complex value for $\boldsymbol{a}$ which relates to the amplitude and phase of $\boldsymbol{M}^{int}$, not simply the projection of the induced-dipole onto the applied field direction. In this context, the scalar product in Eq. 1 is between the induced dipole and the axis of the applied field rather than the oscillating field direction.

We determine the complex valued $a(\omega)$ using Eq. 2, the $\alpha$ and $n$ values obtained previously (see Fig.1 and Fig.2), and the measured values for the complex dielectric function of the pure solvent. The amplitude and phase of $\boldsymbol{a}$ are plotted for both the fibrils and native lysozyme in Figure 3 (the lines represent different THz frequencies while the points correspond to the average values).

For native lysozyme the average amplitude of $\boldsymbol{a}$, $|\boldsymbol{a}|$, is initially $1.4 \pm 0.3$ and the phase, $\Phi(\boldsymbol{a})$, is $-\pi$. This corresponds to an induced field, $\boldsymbol{M}^{int}$, oscillating in phase with the applied field and having a magnitude consistent with previous reports on a variety of different macromolecules in water solutions (see e.g. Ref.[24] and references therein). A subsequent rapid drop of $|\boldsymbol{a}|$ as the concentration is increased takes the amplitude down to $0.6 \pm 0.1$, where it remains flat for the rest of the concentration range. From this point on, however, the average phase increases from $-\pi$ at the lowest concentration to $-0.4\pi$ at the volume fraction of 3.6%.

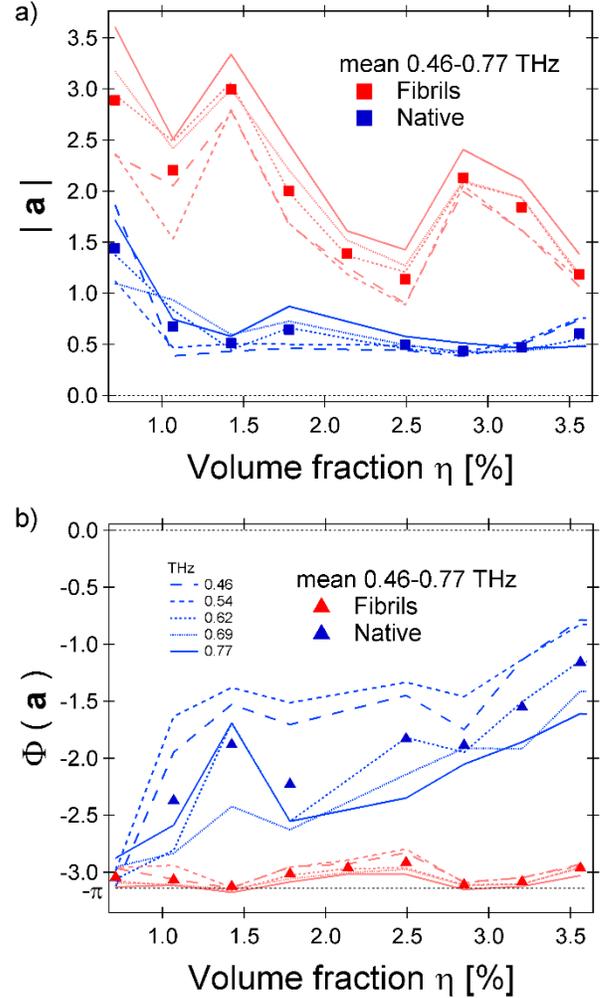

Figure 3. a) Relative amplitude and b) phase of the induced dipole at the protein-solvent boundary versus volume fraction of lysozyme for solutions containing native human lysozyme (blue) and fibrils (red). Squares represent the average response in the 0.46-0.77 THz range, each thin line corresponds to a different frequency from 0.46 THz to 0.77 THz by ~0.07 THz steps (legend inside panel b).

Precise quantitative details of the induced dipole moment at the low concentration limit are determined by the nanoscopic interactions within and between the protein/s and the surrounding water molecules, and requires detailed molecular modelling of the dynamic response, which is beyond the scope of this work. However, as a phenomenological explanation of our results we propose that the two regimes observed correspond to the two cases shown in Figure 4(b). At low concentrations $|\boldsymbol{a}|$ reduces as the probability of the solvation shells overlapping slightly increases, as proposed by Heyden et al[24]. In this case, most of the proteins and their solvation waters are not interacting with the solvation shells of other proteins. Above the critical concentration, where $|\boldsymbol{a}|$ remains constant and Φ(**a**) changes, there is significant overlap between the solvation shells of neighboring proteins. In this regime the solvation water is the dominant "type" of water and the strength of the interaction between the solvation waters and the proteins increases as the average distance between them decreases. This then affects the ability of the induced dipole to follow the field, which results in the phase shift observed.

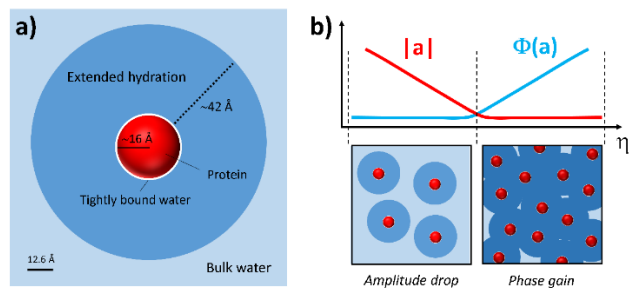

Figure 4. a) Cartoon of a human lysozyme protein (red sphere) in water together with the populations of: i) water molecules tightly-bound to the protein surface (white), ii) extended hydration layers (blue), and iii) unperturbed bulk water (light blue). b) Sketch of the evolution of the modulus and the phase of the induced dipole in a unit volume of solution versus protein concentration. The effect of phase gain at larger concentrations is pictorially represented by darker colors on the bottom right panel.

Simply assuming that all bulk water has been replaced at the point where the trends of both |**a**| and Φ(**a**) change, (i.e. at η ~ 1% or 15 mg/ml for native solutions), and considering a uniform cube lattice of proteins, we estimate[48] an upper limit for the extent of the hydration layer around each native protein. This indicates the hydration layer may extend up to 42 Å from the protein surface for dilute solutions. This value is on the same order of magnitude as previous reports[26-29], and is very similar to the size of the extended hydration layer calculated by molecular dynamics calculations for other protein systems[24].

For the solutions containing fibrils, the plots in Figure 3 for the amplitude (red squares) and phase (red triangles) of **a** are quite different: the amplitude slowly decreases while the phase remains flat at −π. This behavior of the fibrils up to η ~ 3.6% (50 mg/ml) is equivalent to the behavior of native lysozyme for η < 1% (< 15 mg/ml) and is consistent with the phenomenological model proposed above, whereby the dominant trend is determined by the average distance between macromolecular assemblies. For a given concentration of native lysozyme the separation between fibrils, each originating from many closely spaced lysozyme molecules, is much larger, less hydration water is present, and the hydration layers do not overlap. A precise quantitative comparison, however, is prevented by the large heterogeneity in the fibril size (see Methods in SI).

The other major difference observed in the fibrils is that the initial value for |**a**| is 2.8 ± 0.6, significantly larger than the initial value for native lysozyme. This indicates that the induced dipole, $M^{int}$, for the fibrils is larger than for the native. Previous work has revealed that in fibrils the lysozyme tertiary structure changes, such that there is a high density of β-sheets on their surface[51]. This may explain the greater value of $M^{int}$ because the higher surface charge density from the β-sheets can cause a greater disruption to the hydrogen-bonding network of water, however, detailed modelling is required to quantitatively compare the two.

Regardless, the demonstrated ability to reveal differences in the THz response of native proteins and fibrils indicates that terahertz techniques can help to understand the nature of protein-environment interactions and the role they play in stabilizing tertiary structure and driving changes to it[52].

In conclusion, determination of the complex dielectric function of proteins in solution, enabled by time domain THz spectroscopy, allows detailed analysis beyond what is possible from simple absorption measurements. Following this approach we are able to determine both the phase and amplitude of the induced dipole in the volume containing the protein and solvating water. Separating the amplitude and phase of this response reveals that not only can the amplitude of the induced dipole vary as the concentration is changed, but above a concentration threshold so too does the phase of the induced dipole oscillations. While the precise details of the THz response are determined by dynamic atomistic interactions, we propose a phenomenological model that explains the available data, whereby the phase of the induced dipole in the protein-solvent interaction region begins to vary when there is significant overlap between solvation layers of neighboring proteins. This result suggests that indirect electromagnetic protein-protein interactions could take place if mediated by the extended hydration layers surrounding each protein which might also be of great relevance for chemical recognition[52,53].

## AUTHOR INFORMATION NOTES

**Corresponding Author**

* JDavis@swin.edu.au

## ACKNOWLEDGMENT

We thank the Australian Research Council for funding (DP130101690). FN acknowledges funding from Swinburne University via the Early/Interrupted research career scheme 2014. This work has been partially done at the Australian Synchrotron during beamtime AS153/FIR/9692 at the FAR-IR beamline.